\documentclass[fleqn,usenatbib]{mnras}

\usepackage{mathptmx}
\usepackage[T1]{fontenc}
\usepackage{ae,aecompl}
\usepackage{float}
\usepackage{graphicx}	
\usepackage{amsmath}	
\usepackage{amssymb}
\makeatletter

\newcommand{\Rmnum}[1]{\expandafter\@slowromancap\romannumeral #1@}
\makeatother
\newcommand{\A}{\mbox{\normalfont\\A}}

\title[Super Li-rich red clump stars]{ Spectroscopic study of two new super Li rich red clump K giants}

\author[Singh et al.]{
Raghubar Singh$^{1}$\thanks{E-mail: raghubar.singh@iiap.res.in}
Bacham E. Reddy$^{1}$
and
Yerra Bharat Kumar$^{2,3}$
\\
$^{1}$Indian Institute of Astrophysics, Koramangala, Bangalore, 563004, India\\
$^{2}$ Key Laboratory of Optical Astronomy, National Astronomical Observatories, Chinese Academy of
Sciences, Beijing 100012, China\\
$^{3}$ LAMOST Fellow
}

\date{Accepted XXX. Received YYY; in original form ZZZ}

\pubyear{2018}

\begin{document}
\label{firstpage}
\pagerange{\pageref{firstpage}--\pageref{lastpage}}
\maketitle

\begin{abstract}

In this paper,  we report the discovery of two new super Li-rich K giants: HD~24960 and TYC~1751-1713-1. Based on high resolution ($R\approx 60,000$) spectroscopy, we have derived Li abundance of A(Li)$\approx$4.0~dex for both stars. Other elemental abundances are normal of typical K giants. Low ratios of [C/N] $\leq$ $-$ 0.25 and $^{12}C/^{13}C \leq10$ suggest that the stars are in upper RGB phase. Further, based on Gaia astrometry and secondary calibrations using Kepler asteroseismic and LAMOST spectroscopic data, we argue that both stars belong to red clump (RC) phase with core He-burning. Results add to half a dozen already known red clump Li-rich K giants, and support the growing evidence that the origin of Li excess in RC giants seems to be associated with either He-flash at the tip of the red giant branch (RGB) or recent planet/ brown dwarf merger events closer to the RGB tip. 
  
\end{abstract}

\begin{keywords}stars:low mass\,--\,stars: red giants\,--\,stars: red clump\,--\,stars: Li-rich\,--\,stars: atmospheres\,--\,stars: interiors
\end{keywords}

 \section{Introduction}
 Standard stellar models \citep{Iben1967a,Iben1967b} predict 
significant changes, as a result of 1st dredge-up on red giant branch (RGB), to photospheric abundances of some key isotopes of RGB stars: decrease in Li, $^{12}C$ and increase in $^{13}C$ and $^{14}N$ \citep{Charbonnel2005}. 
Models predict ratio of $^{12}C/^{13}C$ in the range of 20 to 25 \citep{Charbonnel1994} from their main sequence value of about 89 \citep{Asplund2009}. Similarly, Li is predicted to be a factor of about 30-70 less in RGB stars after the 1st dredge-up compared to their progenitors on the main sequence (\citealt{Iben1967a}, \citealt{Lagarde2012}). However, standard models do not predict further changes to abundances beyond the 1st dredge-up \citep{Iben1967a}.

In general, 1st dredge-up predictions agree well with the observations of RGB stars below the luminosity bump. However, giants at or above the bump, contrary to the standard predictions, show significantly changed abundances from the 1st dredge-up (\citealt{Lind2009b}, \citealt{Kirby2016}). The changes are in the same direction as 1st dredge-up but much severe. For example $^{12}C/^{13}C$ drops to less than 10 in most cases, and in some cases as low as its equilibrium value  of 3 to 4 (\citealt{Gilroy1989}, \citealt{Gilroy1991}). 
Similarly, observed Li abundance in upper RGB stars is significantly low with typical value of A(Li) $\leq$ 0.5~dex \citep[e.g.][]{Brown1989} which is about three magnitudes less than the initial maximum main-sequence value of A(Li)$\sim$3.2~dex \citep{Lambert2004}. 
The rapid decrease of carbon isotopes post-1st dredge-up is attributed to some kind of extra-mixing during stellar evolution through luminosity bump \citep{Denissenkov2003}. 
It is not known what exactly causes extra-mixing. However, there are a number of possible mechanisms such as thermohaline mixing \citep{Eggleton2006}, magneto-thermohaline mixing \citep{Denissenkov2009}, rotationally induced mixing \citep{Denissenkov2004} etc. The general understanding is that it happens at the luminosity bump at which H-burning shell advances upward causing the removal of chemical composition discontinuity \citep{Dalsgaard2015}, a barrier to mixing, left behind by the 1st dredge-up. The removal of discontinuity in the abundance profile revives the mixing with the outer convective envelope resulting further reduction in the values of $^{12}C/^{13}C$  and Li abundance. 

Against these general observational trend and theoretical predictions, a small group of low mass RGB stars show large amounts of Li, A(Li) $\geq$ 1.5~dex (e.g \citealt{Kumar2011a}). In some giants enhancements are 2 to 4 orders of magnitude above the expected values, exceeding the initial maximum main sequence or ISM values of 
 A(Li) = 3.2~dex \citep{knauth2003}, which are known as super Li-rich giants \citep[e.g.][]{Kumar2011a}. Presently, there are close to 200 Li-rich K giants  of which about 2 dozen are super Li-rich \citep[e.g.][]{Casey2016, Smiljanic2018}.

In spite of finding a large number of Li-rich stars on RGB, there is no consensus on the origin of Li excess. One of the principal reasons for the continued debate may probably be due to lack of precise stellar evolutionary phase of Li-rich stars on RGB. Though Li-rich stars were reported to be across RGB starting from sub-giant phase (e.g. \citealt{Martell2013, Li2018}) to all the way to red clump with He-core burning phase \citep[e.g.][]{SilvaAguirre2014, kumar2018}, majority of them are known to be  concentrated in a narrow luminosity range of 1.4 $\leq log (L/L_{\odot}) \leq 2.0$ overlapping with the luminosity bump region as well as the red clump region (\citealt{Charbonnel2000}, \citealt{Kumar2011a}). Since bump is associated with stellar internal changes many studies explored to explain the anomalous Li enhancement during the bump as a consequence of extra-mixing \citep[e.g.][]{Palacios2001}. 

However, the recent discovery of Li-rich K giants (e.g. \citealt{SilvaAguirre2014, kumar2018, Smiljanic2018}) which are in  Kepler or CoRoT field,  and determining their evolutionary phase as red clump with He-core burning based on asteroseismic analysis \citep{Bedding2011} shifted the focus of Li enhancement to the He-flash or post He-flash phase.  Precise determination of stellar evolutionary phase of Li-rich stars on RGB  has significant consequences  as Li enhancement scenarios vary with evolutionary phase.

We report here  high resolution spectroscopic analysis of two new super Li-rich K giants: HD~24960 and TYC~1751-1713-1 for which no prior spectroscopic analysis of elemental abundances is reported. Results in this paper will add to
half a dozen known Li-rich K giants of red clump.    

\section{Observations}
\begin{figure}
\centering
\includegraphics[width = \columnwidth]{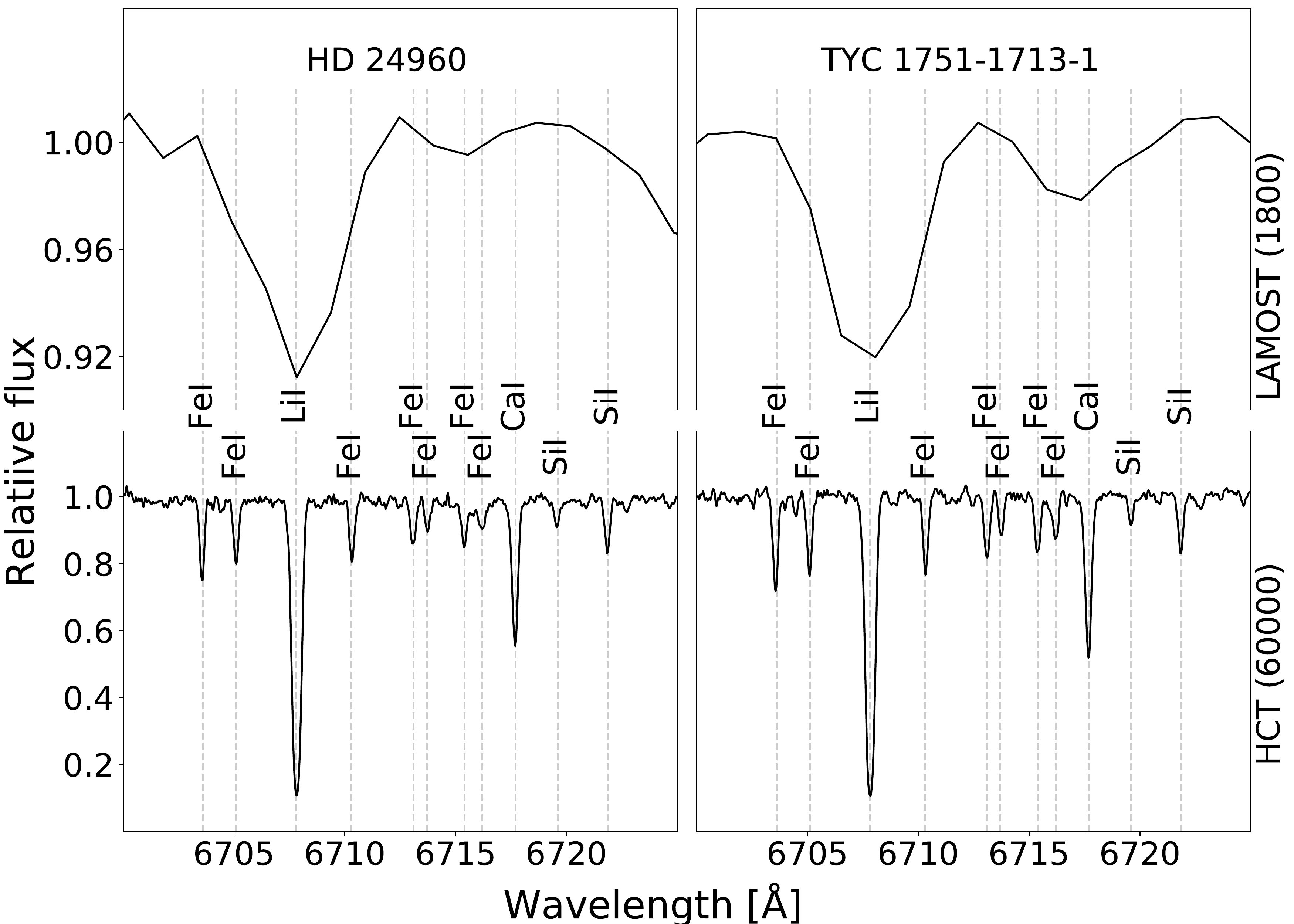}
\caption{ Sample spectra of stars in the Li region:  Low resolution spectra from LAMOST  (top two panels) and the corresponding  high resolution spectra (bottom two panels).}
\label{fig:lamostspectra}
\end{figure}

LAMOST\footnote{http://dr3.lamost.org/} low resolution (R=1800) spectra of the two stars in this study have been screened out from the spectra of much larger sample of K giants which have been scanned for finding new Li-rich K giants. Sample LAMOST spectra of two stars showing Li line at 6707~\AA\ are shown in Fig~\ref{fig:lamostspectra} (top panels). Line at 6707~\AA\ is quite strong and characteristic of unusual Li abundance \citep{kumar2018}. Subsequently, the two candidate Li-rich K giants have been observed with 2.0-m Himalayan Chandra Telescope (HCT) equipped with the high resolution spectrograph or HESP (Hanle Echelle Spectrograph). Single spectrum of  10 minutes exposure for HD~24960 (m$_{v}$ = 8.00) and three spectra of 30 minutes exposure each  for TYC 1751-1713-1 (m$_{v}$ = 9.42) are obtained. Spectra have good signal-to-noise ratio of SNR $>$ 100.

Standard reduction procedure such as bias, flat field corrections, and extraction of one dimensional spectra of flux versus pixel has been followed using {\it Image Reduction Analysis Facility} $(IRAF)$ tools. For identification and removal of atmospheric lines, we observed HD~5394, a bright  hot star of spectral type BIV with very high rotational velocity of $vsini = 265~Km s^{-1}$ during the same night. Stellar spectra are wavelength calibrated using Th-Ar arc calibration spectra obtained immediately after each star's observation. Finally spectra were continuum fitted for the measurement of equivalent widths (EWs). Cross correlation of stellar spectra with a template spectrum yields radial velocity ($R_{v}$) of $-$36.1~km $s^{-1}$ and 5.06 km $s^{-1}$ for HD~24960 and TYC~1751-1713-1, respectively. Sample spectra of two stars in the regions of Li line at 6707~\AA\ are shown in Figure~\ref{fig:lamostspectra} (bottom panels). 

\begin{figure}
\centering
\includegraphics[width = \columnwidth]{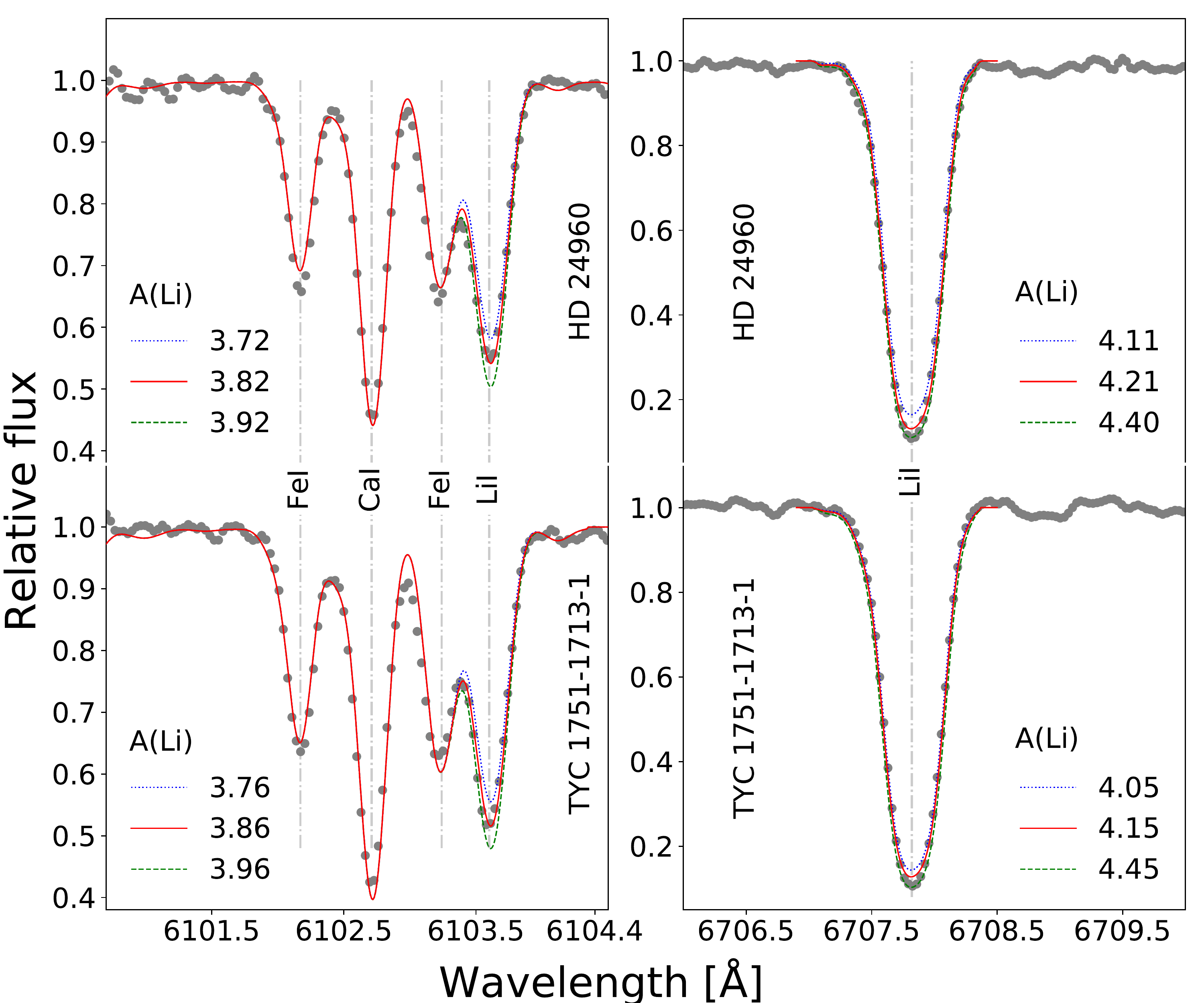}
\caption{Synthetic spectra (continuous line)  around resonance Li line 6707.8 \AA\ and subordinate Li line at 6104 \AA\ for different abundances compared with observed spectra (dotted line).}
\label{fig:lilines}
\end{figure}

\begin{figure}
\centering
\includegraphics[width = \columnwidth]{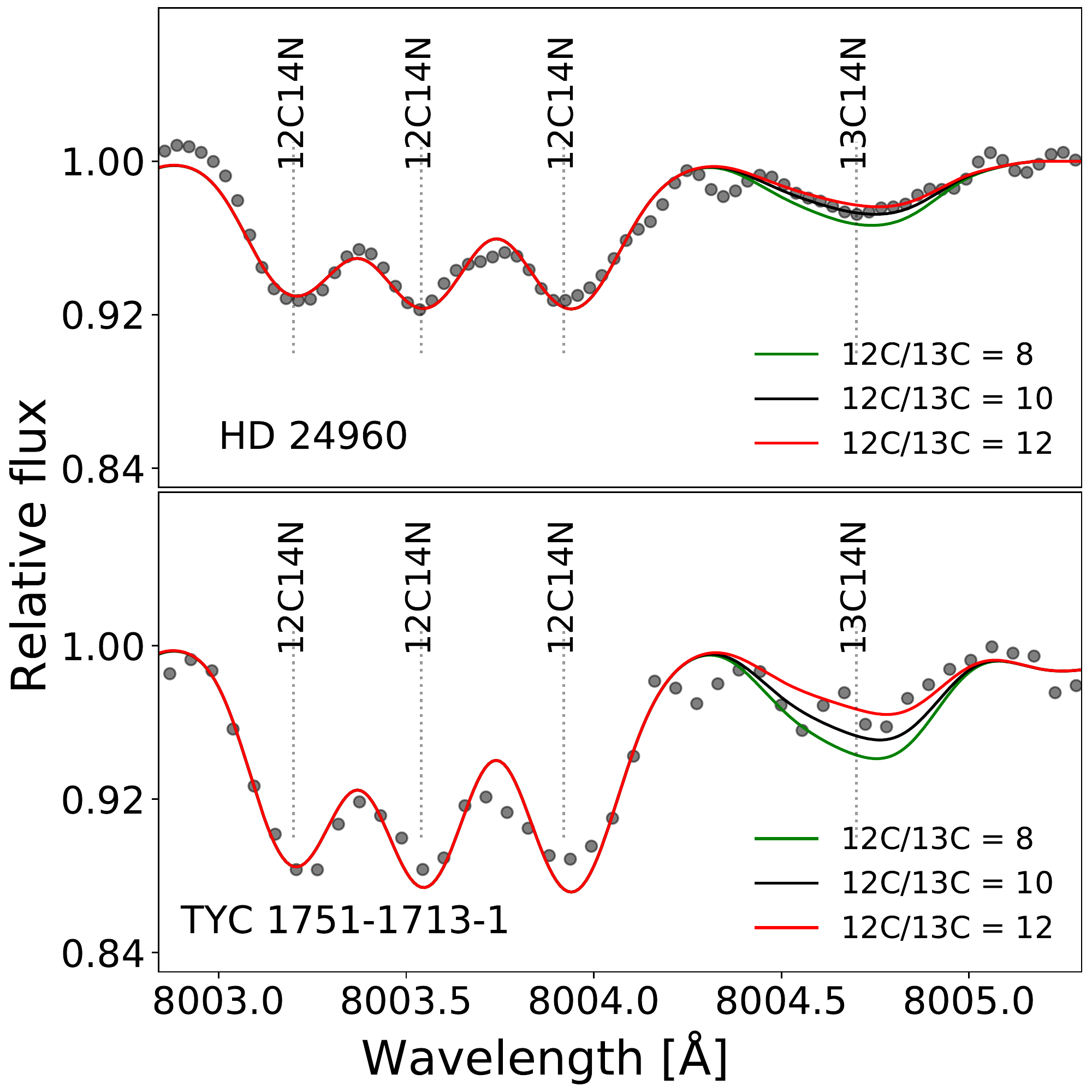}
\caption{Synthesis of $^{13}CN$ spectral line compared with observed spectra (dotted line). }
\label{fig:cnlines}
\end{figure}

\subsection{ Analysis and Results}
Abundances have been obtained for 24 chemical elements. For each element a number of representative lines of moderately strong, free of known blends and having reliable atomic data were identified based on a typical K giant, Arcturus ($\alpha$-Boo) spectral atlas \citep{Hinkle2000}. Line strengths i.e equivalent widths (EWs) are measured from continuum-normalized spectra using $IRAF$ tools. Depending on line strengths and possible blends such as Hyper fine structure (HFS), abundances are derived either by matching observed spectra with the synthetic spectra or matching individual line EWs with the predicted EWs in case of clean lines with free of blends. For the analysis, we used LTE stellar atmospheric models from Kurucz \citep{CastelliKuruxz2004}, and the recently updated radiative transfer code MOOG 2013 version \citep{Sneden1973}. 

\subsubsection {Atmospheric parameters}

Atmospheric parameters: $\mathrm{T_{eff}}$, log $g$, microturbulance velocity ($\xi_{t}$) and [Fe/H] have been derived using a set of well defined \ion{Fe}{i} and \ion{Fe}{ii} lines taken from the compilation of \citet{Reddy2003} and \citet{Ramirez2011}. A total of 40 \ion{Fe}{i} and 10 \ion{Fe}{ii} lines with EWs less than about 110~m\AA\ are used. Standard iterative procedure is adapted for deriving $\mathrm{T_{eff}}$, log $g$ and $\xi_{t}$. We chose $ \mathrm{T_{eff}}$ and $\xi_{t}$ of an atmospheric model for which \ion{Fe}{i} lines' abundances are independent of their corresponding lower excitation potential (LEP) and their EWs, respectively.  In case of log $g$,  models are varied until neutral average \ion{Fe}{i} abundance is equal to that of singly ionized \ion{Fe}{ii} abundance. Derived parameters along with the uncertainties are given in Table~\ref{tab:stellar parameter}. Uncertainties are estimated by using sensitivity of abundance trends to changes in small steps to respective parameters. We find no appreciable effect on lines' abundance versus LEP relation for a change up to  $\delta \mathrm{T_{eff}}$ = $\pm$ 70~K for TYC~1751-1713-1 and 35~K for HD~24960. Thus, $\pm$ 70~K and 35~K is taken as uncertainty in the  determination of $\mathrm{T_{eff}}$. 
Also, values of $\mathrm{T_{ eff}}$ and log $g$  derived using photometry combined with Gaia parallaxes \citep{Gaia2018}. Photometry is taken from from 2MASS \citep{Skrutskie2006} and Simbad\footnote{http://simbad.u-strasbg.fr/simbad/sim-fbasic}. The de-reddened colors of (V-K)$_{\odot}$ and (J-K)$_{\odot}$ combined with empirical relations from \citet{HernandezBonifacio2009} yield an average $\mathrm{T_{ eff}}$ of 4870$\pm$60~K for TYC 1751-1713-1 and {4800$\pm$80~K} for HD~24960. Reddening value of E(B-V) = 0.09 is taken from dust maps \citep{Green2015} in the direction of stellar sources. Values of log $g$ are also derived using a relation; 

$$ \log g = \log g_{\odot} + \log \Big( \frac{M}{M_{\odot}}\Big) - \log \Big( \frac{L}{L_{\odot}}\Big) + 4 \times \log \Big( \frac{T}{T_{\odot}}\Big) $$

Masses are estimated using stellar evolutionary tracks \citep{Paxton2018} with corresponding stellar metallicity  that pass through respective positions of Luminosity and $\mathrm{T_{eff}}$ in the HR-diagram. Derived $\mathrm{T_{eff}}$ and log $g$ values using astrometry and photometry along with  spectroscopic values are given in Table~\ref{tab:stellar parameter}. Values are in good agreement within the quoted uncertainties. We adopt values from spectroscopic analysis.

\begin{table}
\caption{Derived stellar parameters of program stars.}
\label{tab:stellar parameter}
\centering
\begin{tabular}{p{2.8cm} p{2.3cm} p{1.8cm}}
\hline
  & TYC~1751-1713-1 & HD~24960 \\ 
\hline
RA                       & 01:21:41.845         & 03:59:27.632         \\ [1ex]
DEC                      & +25:15:06.79         & +36:19:06.53         \\ [1ex]
V (mag)                  & 9.42                 & 8.00                 \\ [1ex]
HESP Spectra SNR         & 110                  & 120                  \\ [1ex]
Parallax (mas)           & $1.95  \pm 0.04 $    & $ 3.67 \pm0.04  $    \\ [1ex]     
$\log g (gm s^{-2})$     & $ 2.58 \pm 0.17 $    & $ 2.39 \pm 0.19 $    \\ [1ex]
logg$_{photometric}$     & $ 2.45 \pm 0.04 $    & $ 2.40 \pm 0.03 $    \\ [1ex]
Teff [K]                 & $ 4830 \pm 70   $    & $ 4835 \pm 35   $    \\ [1ex]
Teff$_{photometric}$     & $ 4870 \pm 60   $    & $ 4800 \pm 80   $    \\ [1ex] 
$\xi_{t}$($km s^{-1}$)   & $ 1.47 \pm 0.15 $    & $ 1.41 \pm 0.04 $    \\ [1ex]
[Fe/H] (dex)             & $-0.25 \pm 0.10 $    & $ -0.45 \pm 0.04 $   \\ [1ex]
vsini ($km s^{-1} $)     & $ 4.40 \pm 0.8  $    & $ 6.0  \pm 1    $    \\ [1ex]
$V_{m}$ ($km s^{-1}$)    & $ 3.60 \pm 0.10 $    & $ 4.0  \pm 0.20 $    \\ [1ex]
[C/Fe]                   & $ 0.06 \pm 0.20 $    & $ 0.21 \pm 0.09 $    \\ [1ex]
[N/Fe]                   & $ 0.37 \pm 0.16 $    & $ 0.94 \pm 0.18 $    \\ [1ex]
[O/Fe]                   & $ 0.11 \pm 0.17 $    & $ 0.33 \pm 0.12 $    \\ [1ex]
$^{12}$C/$^{13}$C        & $ 10  \pm 1     $    & $ 10 \pm 2 $         \\ [1ex]
A(Li)$_{LTE,6103}$ \AA   & $ 3.86 \pm 0.13 $    & $ 3.82 \pm 0.07 $    \\ [1ex]
A(Li)$_{NLTE,6103 }$ \AA & $ 3.97 \pm 0.13 $    & $ 3.94 \pm 0.07 $    \\ [1ex]
A(Li)$_{LTE,6708 }$ \AA  & $ 4.15 \pm 0.15 $    & $ 4.21 \pm 0.10 $    \\ [1ex]
A(Li)$_{NLTE,6708 }$ \AA & $ 4.05 \pm 0.15 $    & $ 4.06 \pm 0.10 $    \\ [1ex]
Age (Gyr)                & 8.8                  & 8.1                  \\ [1ex]
Mass ($M_{\odot}$)       & $ 1.3 \pm 0.05 $     & $ 1.2 \pm 0.05 $     \\ [1ex]
Sptype                   & K2III                & K2III                \\ [1ex]

\hline
\end{tabular}
\end{table}

\subsubsection{Abundances}
{\bf {Lithium}}: Li abundance has been derived from resonance line at 6707.8~\AA\ and a subordinate line at  6103.6~\AA. In general, later transition is not seen unless stars have extremely large Li abundance. Abundance are derived using spectral synthesis to account for weak blends, and hyperfine structure (HFS) in case of resonance line at 6707~\AA. For line synthesis, we adopted line list along with oscillator strengths from the compilation of \citet{Reddy2002} and  HFS from \citet{Hobbs1999}. Spectral synthesis of respective lines yields Li abundances of A(Li) = 4.15\footnote{A(Li) = log(N(Li)/N(H)} and 3.86~dex for TYC~1751-1713-1 and  A(Li) = 4.18 and 3.82~dex for HD~24960. Comparison of observed spectra with the predicted spectra for different Li abundances for both stars are shown in Fig~\ref{fig:lilines}.  Both transitions are affected by non-LTE effects \citep{Lind2009a} and the derived abundances are corrected using their recipe. The average non-LTE corrected abundances are: A(Li) = 4.0~dex for TYC~1751-1713-1 and 4.01~dex for HD~24960 ( See Table~\ref{tab:stellar parameter}). 

\noindent
{{\bf{C,N,O \& $^{12}C/^{13}C$ }}}: Abundance ratios of C/N and, in particular, isotopic ratios of $^{12}C/^{13}C$ are key tracers of evolution on RGB. We have derived carbon abundance using three well defined atomic lines at 5052.15~\AA, 5380.36~\AA\ and 6587.62~\AA\ with $gf$ values taken from \citep{Wiese1996}. Atomic N lines in the spectra are quite weak as the stars are too cool and the N transitions in the visible spectra are of high voltage lines.  For N abundance, we relied on a few selected CN lines in the region of 8000-8006~\AA\ and with the derived C abundance as input. Basic line data of CN lines for spectral synthesis is taken from \citet{deLaverny1998}.
Oxygen abundances are based on two forbidden [OI] lines at 6300.30~\AA\ and 6363.78~\AA. In case of 6300.30~\AA\ synthesis, we included line list  with atomic data given in \citep{AllendePrieto2001}.  Values for isotopic ratio $^{12}C/^{13}C$ have been derived using a molecular line of $^{13}C^{14}N$ at 8004.6~\AA. Derived C, N,O and $^{12}C/^{13}C$ ratios are given in Table~\ref{tab:stellar parameter}. \\

\begin{table*}
\vspace{1cm}
\caption{ Summary of elemental abundances with estimated errors of the two program stars.}
\label{tab: elements}
\centering
\begin{tabular}{ p{2cm} p{2cm} p{2cm} p{2cm} p{2cm} p{2cm} p{2cm} }  \hline  
\multicolumn{1}{l }{Element} & \multicolumn{3}{c }{HD~24960 ([Fe/H]=-0.45)} & \multicolumn{3}{c}{TYC~1751-1713-1 ([Fe/H]=-0.25)}\\ [1ex]
    & A(X)  & [X/Fe] & $\sigma _{[X/Fe]}$ & A(X) & [X/Fe] & $\sigma _{[X/Fe]}$ \\ [1.5ex]
          (1)  &  (2)  &  (3)   & (4)   &  (5)  & (6)     &  (7)  \\  \hline
          Li   &  4.00 &   ...  &  ...  &  4.01 &   ...   &  ...  \\
          C    &  8.20 &   0.21 &  0.09 &  8.24 &    0.06 &  0.20  \\
          N    &  8.33 &   0.94 &  0.18 &  7.95 &    0.37 &  0.16  \\
 $\mathrm{[O]}$&  8.58 &   0.33 &  0.12 &  8.55 &    0.11 &  0.17 \\
  \ion{Na}{i}  &  5.95 &   0.15 &  0.08 &  6.12 &    0.13 &  0.14 \\
  \ion{Mg}{i}  &  7.19 &   0.03 &  0.08 &  7.29 &   -0.06 &  0.13 \\
  \ion{Al}{i}  &  6.05 &   0.04 &  0.08 &  6.30 &    0.10 &  0.13 \\
  \ion{Si}{i}  &  6.99 &  -0.08 &  0.08 &  7.17 &   -0.09 &  0.14 \\
  \ion{Ca}{i}  &  5.94 &   0.04 &  0.09 &  6.16 &    0.07 &  0.15 \\
  \ion{Sc}{ii} &  2.70 &  -0.01 &  0.11 &  3.05 &    0.15 &  0.16 \\
  \ion{Ti}{ii} &  4.46 &  -0.05 &  0.10 &  4.73 &    0.03 &  0.17 \\
  \ion{V}{ii}  &  3.33 &  -0.16 &  0.10 &  3.70 &    0.02 &  0.16 \\
  \ion{Cr}{i}  &  5.22 &   0.02 &  0.08 &  5.35 &   -0.04 &  0.16 \\
  \ion{Fe}{i}  &  7.01 &   0.00 &  0.09 &  7.20 &    0.00 &  0.17\\
  \ion{Co}{i}  &  4.46 &  -0.09 &  0.11 &  4.87 &    0.13 &  0.15 \\
  \ion{Ni}{i}  &  5.84 &   0.06 &  0.09 &  5.98 &    0.01 &  0.17 \\
  \ion{Cu}{i}  &  3.85 &   0.10 &  0.12 &  4.28 &    0.34 &  0.18 \\
  \ion{Zn}{i}  &  4.15 &   0.03 &  0.14 &  4.29 &   -0.02 &  0.18 \\
   \ion{Y}{i}  &  1.53 &  -0.24 &  0.12 &  1.98 &    0.02 &  0.15 \\
  \ion{Zr}{i}  &  2.16 &   0.02 &  0.11 &  2.38 &    0.05 &  0.18 \\
  \ion{Ba}{ii} &  2.02 &   0.28 &  0.10 &  2.27 &    0.34 &  0.20 \\
  \ion{La}{ii} &  0.87 &   0.18 &  0.12 &  1.07 &    0.19 &  0.15 \\
  \ion{Ce}{ii} &  1.24 &   0.10 &  0.14 &  1.35 &    0.02 &  0.19 \\
  \ion{Nd}{ii} &  0.96 &  -0.05 &  0.17 &  1.49 &    0.29 &  0.22 \\
\hline 
\end{tabular}
\end{table*}

\noindent
{\bf {Other Elements}}:  
In addition to key elements like Li, C, N and  $^{12}$C/$^{13}$C, we have derived abundances of other elements to look for any possible trends with Li excess. Atomic data for most of the transitions are taken from the compilation of \citet{Reddy2003} and \citet{Ramirez2011}. The derived average abundances of elements from Li to Nd for which we could measure reliable EWs are tabulated in Table~\ref{tab: elements}. Except the abundance of Li,  other elemental abundances as defined by  [X/Fe] ($= [X/H]_{star} - [Fe/H]_{star}$) do not suggest any abnormality, and they are typical for stars on the RGB. Also, abundance patterns are typical of Galactic thin disk component to which both stars belong as per the kinematic measurements with probability of being thin disk by more than 95$\%$ \citep{Reddy2006}. \\

\noindent
{\bf {Uncertainties}}: 
Quoted errors in average abundances given in Table~\ref{tab:stellar parameter} \& \ref{tab: elements}  are quadratic sum of uncertainties in respective abundances due to estimated uncertainties in model parameters: $\Delta T_{\rm eff}$, $\Delta log $, $\Delta [Fe/H]$ and $\Delta \xi_{t}$. Also included is the estimated uncertainty in measuring EWs ($\Delta EW$) which is a function of S/N ratio of spectra \citep{cayrel1988}. The individual uncertainties in each of these parameters are assumed to be independent,  and the net error  in each abundance is given by
$$  X_{error} = \sqrt{\Delta Teff^{2} + \Delta logg ^{2} + \Delta Vt^{2} + \Delta [Fe/H]^2 + \Delta EW ^{2}}$$

Obviously, abundances from neutral lines are more sensitive to $\Delta T_{\rm eff}$, ionized species (\ion{Fe}{ii}, \ion{Ba}{ii} etc.) to $\Delta log g$, and the abundances derived using stronger lines are more sensitive to $\Delta \xi_{t}$. Effect of $\Delta [Fe/H]$ is very little or nil on abundances. 
\noindent
\subsubsection{$Vsini$ and infrared excess}
Many studies speculated on the possible correlation between Li-excess, and $vsini$ and IR excess which can be one of the key evidences for  planet/sub-dwarf merger scenarios \citep{Carlberg2010}. \\  
\noindent

In general, average stellar rotational velocities ($vsini$) for K giants are low,  at about 2 Km s$^{-1}$, compared to their progenitors on the main sequence \citep{deMadeiros1996}. As stars evolve off to RGB their  $vsini$
drops significantly due to angular momentum conservation. However, a few Li-rich giants do show higher $vsini$ values than the typically observed values prompting speculations of correlation between $vsini$ and Li-excess \citep{Drake2002}. In the case of two stars in this study, we have derived $vsini$ values using two neutral \ion{Fe}{i} lines 6703.5~\AA\  and 6705.1~\AA. Observed spectral profiles are compared with the profiles predicted for given input atomic data, abundance, and instrumental broadening derived from Th-Ar calibration spectrum. By varying values of $vsini$ and macro turbulence($V_{m}$) simultaneously, we found best fits with minimum $\chi^{2}$ value for $vsini$= 6$\pm$1 km s$^{-1}$ and $V_{m}$= 4.0$\pm$0.2 km s$^{-1}$ for HD~24960, and 4.4$\pm$0.8 km s$^{-1}$ and $V_{m}$= 3.6$\pm$0.1 km s$^{-1}$ for TYC~1751-1713-1. Values of $V_{m}$ and $Vsini$ are given in Table~\ref{tab:stellar parameter}. \\

\begin{figure}
\centering
\includegraphics[width=\columnwidth]{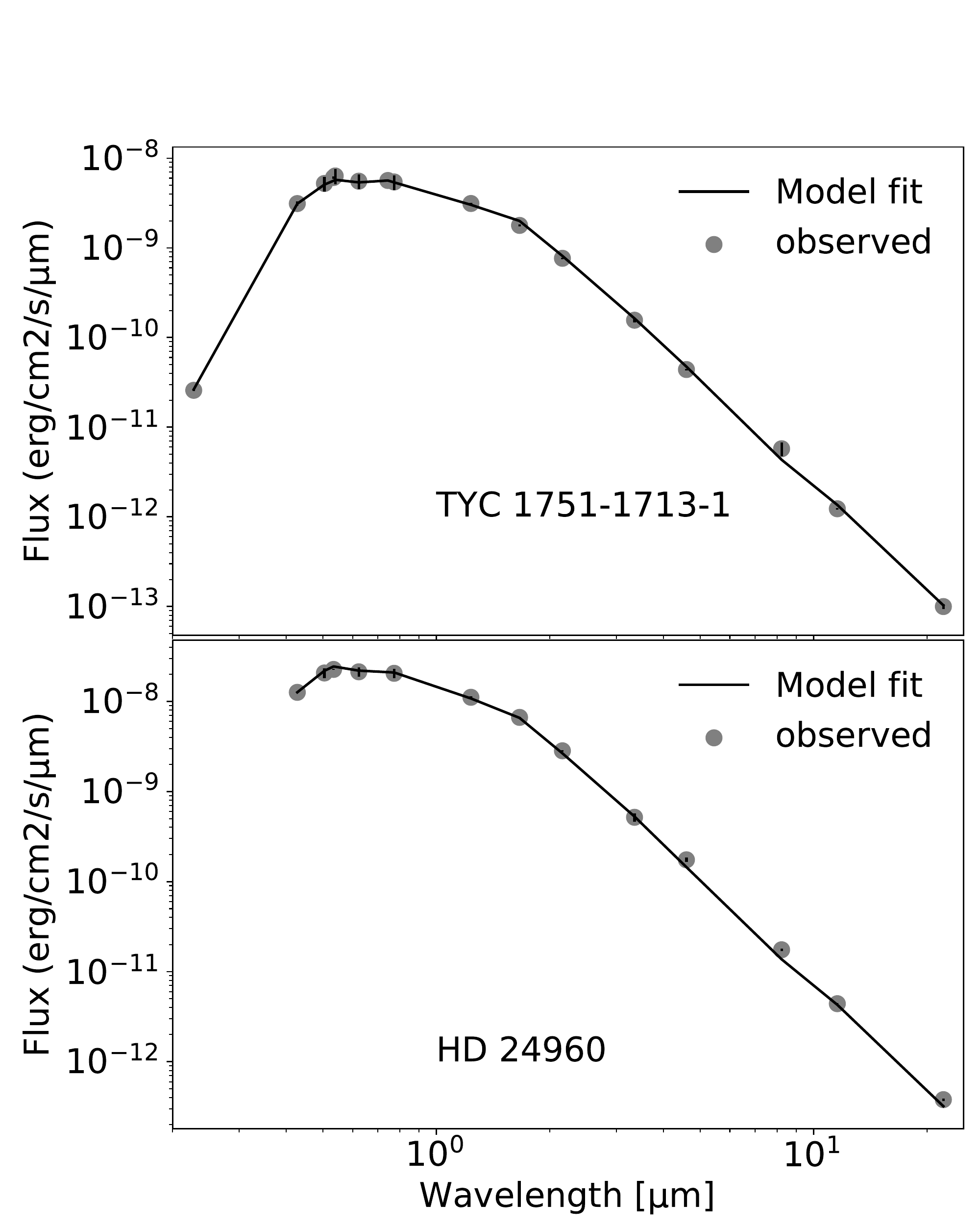}
\caption{ Model SED fits (continuous line) of TYC~1751-1713-1 and HD~24960 with observed fluxes (filled circles).}
\label{fig:sed}
\end{figure}

\begin{figure}
\centering
\includegraphics[width=\columnwidth]{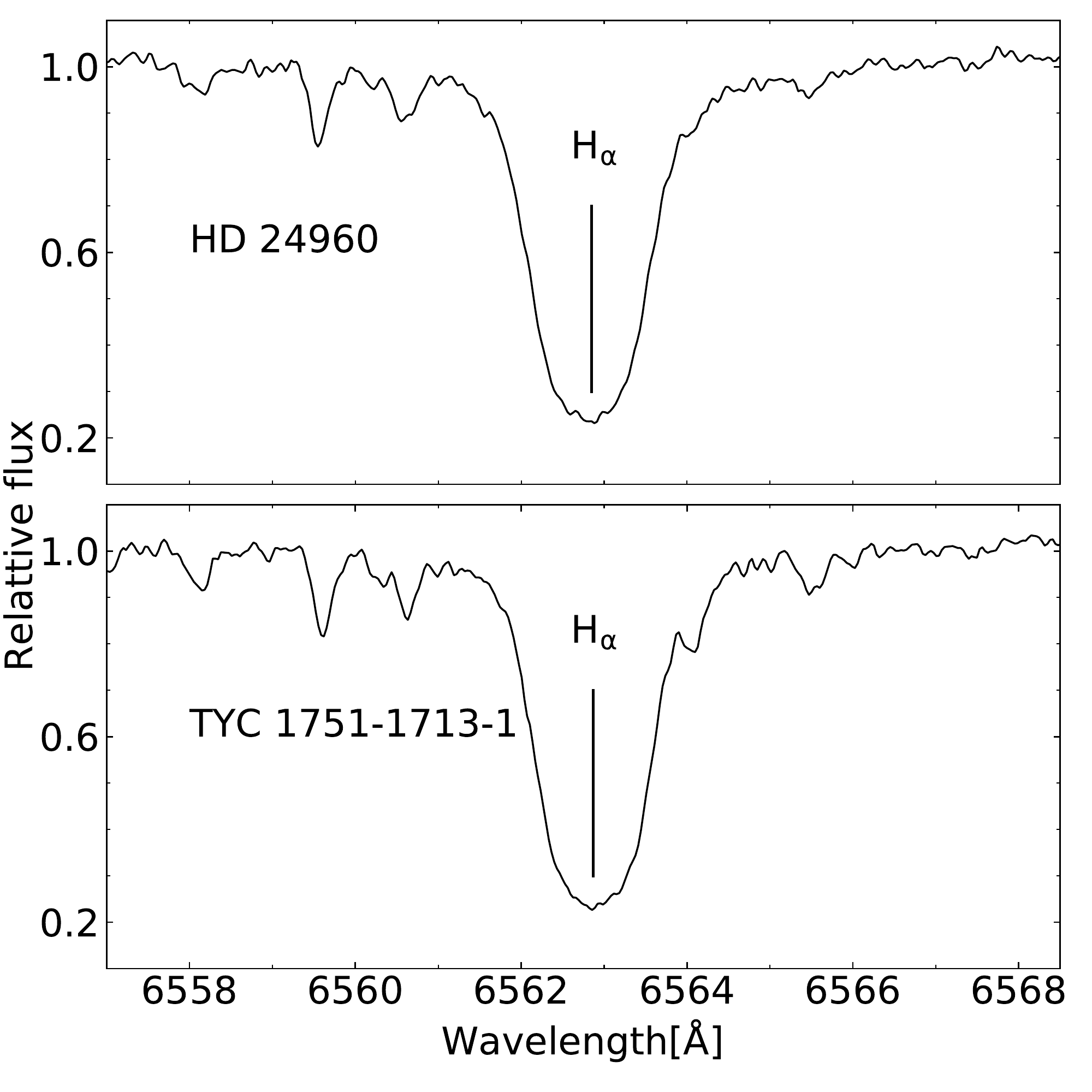}
\caption{ Observed H$_{\alpha}$ of profiles show no asymmetry. Central line is a bisector of profiles.}
\label{fig:halpha}
\end{figure}

\noindent
For determining IR excess, we used spectral energy distribution (SED). The observed available fluxes in different bands are fitted with theoretical Kurucz atmospheric flux models \citep{Castelli1997} for a set of derived atmospheric parameters. We used mid IR fluxes from AKARI \citep{Ishihara2010} and WISE \citep{Wright2010}, and near IR fluxes from 2MASS \citep{Skrutskie2006} survey. The data for optical bands (G, G$_{BP}$, G$_{RP}$) comes from Gaia survey \citep{Gaia2018}. Near- and mid-IR fluxes of both stars are of very good quality as indicated by flags in respective catalogues. As shown in Figure~\ref{fig:sed} neither of the stars show evidence of IR excess at least in the near- and mid-IR regime. Stars may have excess in far IR due to cold dust for which we don't have data because far IR fluxes of 24$\mu$, 60$\mu$ and 100$\mu$ given in IRAS catalogue \citep{Moshir1990} are upper limits which are not considered in fitting SEDs. Lack of IR excess, at least in the near IR, suggests lack of mass loss activity either at present or in the recent past. This is further corroborated by the near symmetric $H_\alpha$ profiles \citep{Meszaros2009} (see Figure~\ref{fig:halpha}) as asymmetric profiles are commonly used for
stellar mass-loss. \\  
 
\begin{figure}
\centering
\includegraphics[width = \columnwidth]{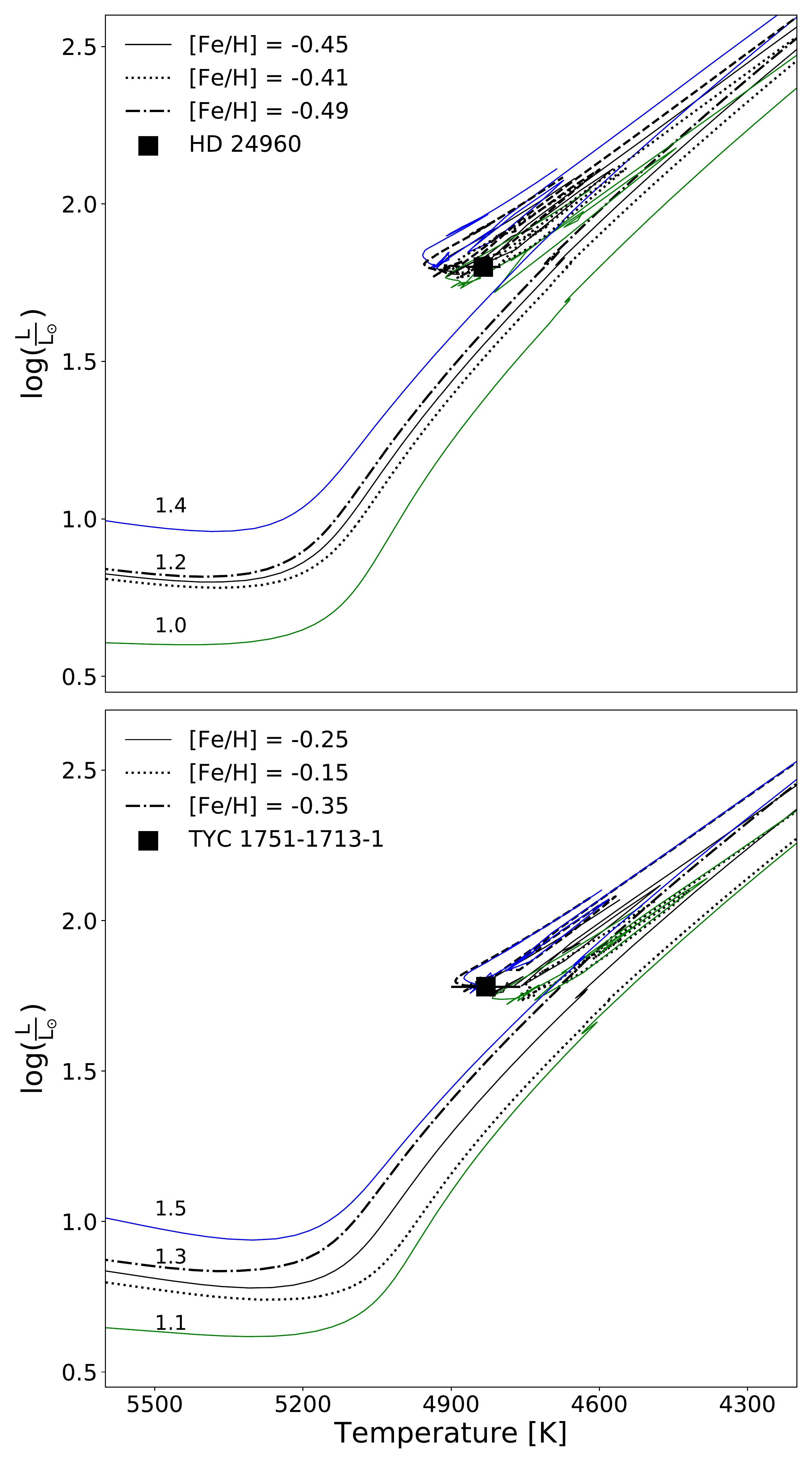}
\caption{ Location of two super Li-rich stars (filled squares) in the HR diagram. Shown are the $ MESA $ evolutionary tracks \citet{Paxton2011} for masses 1.2, 1.3 M$_{\odot}$ with corresponding metallicities.}  

\label{fig:hrd}
\end{figure}

\subsection {Evolutionary Status} Evolutionary phase of Li-rich stars is one of the key issues which is not unambiguously determined for most of the Li-rich RGB stars. The resolution of the issue is closely tied to the understanding of Li excess origin in K giants. Here, we used two different methods to determine stellar evolutionary phase. First one involves simply placing stars in the HR diagram of luminosity versus $T_{\rm eff}$ plane and compare with theoretical evolutionary tracks. As shown in Figure~\ref{fig:hrd} evolutionary tracks from MESA \citep{Paxton2011, Paxton2018} with corresponding metallicity suggest that stars are at the red clump region,  post He-flash at the RGB tip. Errors in luminosity due to errors in parallaxes, and $T_{\rm eff}$ are shown as vertical and horizontal bars. Parallaxes are taken from Gaia survey \citep{Gaia2018}. Other source of error in determining star's position in the HR-diagram is the metallicity. We plotted two tracks for each star for upper and lower limits of [Fe/H] (tracks with broken lines). Tracks for given average metallicities of masses  of $1.2 \pm0.05$\ M$_{\odot}$ for HD~24960 and $1.3\pm0.05$\ M$_{\odot}$ for TYC~1751-1713-1 best converge with stars positions in $log( L/L_{\odot})$ - $T_{\rm eff}$ plane. Though the position appears to coincide well with the red clump region,  given the very small gap between the clump and the bump for field stars of metallicities closer to the sun, it would be difficult to rule out the possibility of their location on the RGB either at or close to the luminosity bump in HR diagram. This is one of the main difficulties to separate stars of RC from that of the bump \citep{Kumar2011a}.  
 
\begin{figure}
\centering
\includegraphics[width = \columnwidth]{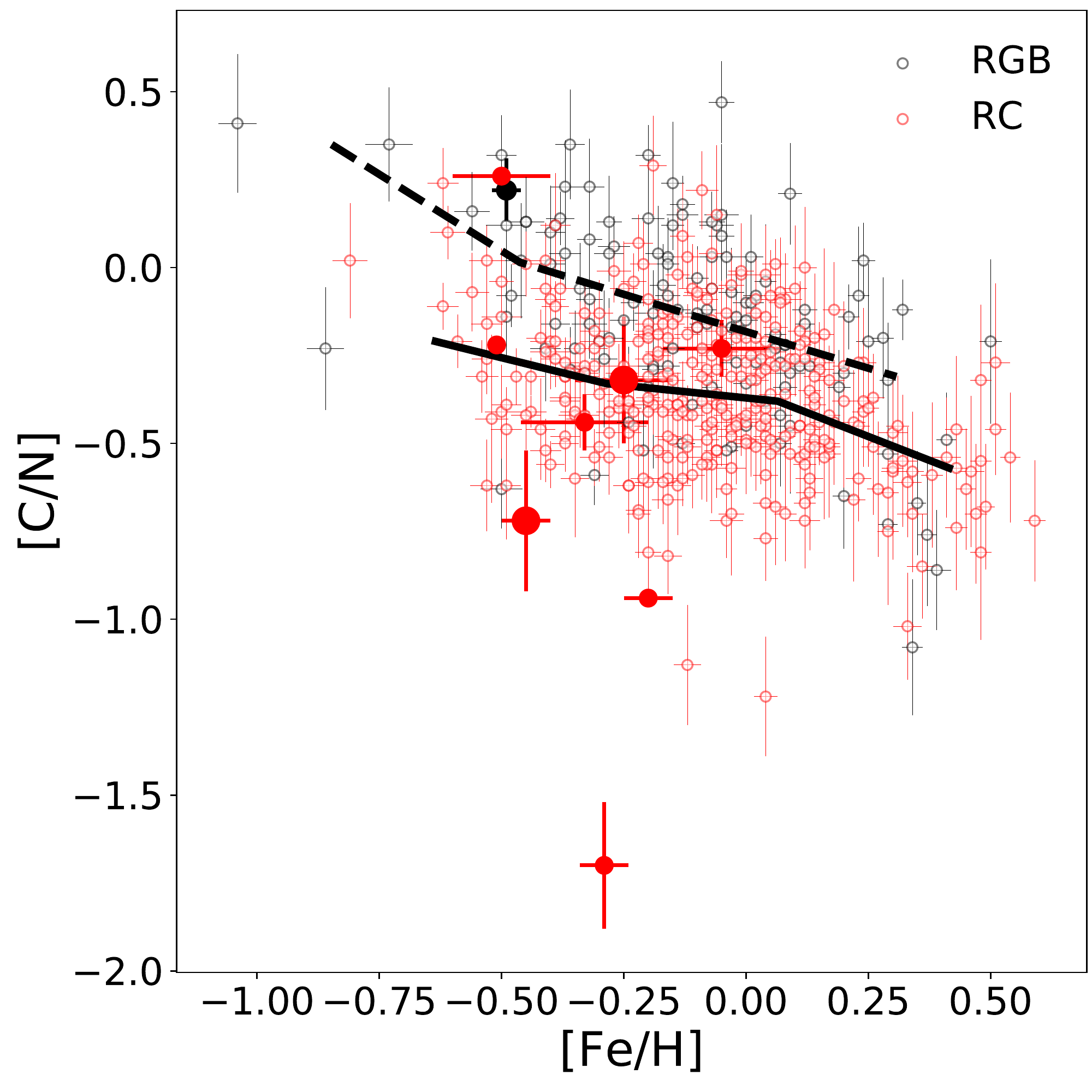}
\caption{ Plot of [C/N] vs [Fe/H] of known normal RC (red open circles) and RGB (black open circles) stars. Li-rich RC stars from Table~4 are shown as red filled circle and the lone Li-rich RGB star KIC~9821622 is shown as filled black circle.  Black solid and black dashed lines are running median of [C/N] abundance of RC and RGB stars, respectively. Note, two stars in this study (large filled red circles) are well within RC regime of the plot.}
\label{fig:cn}
\end{figure}
 
Second method pertains to asteroseismology for differentiating RGB stars with H-shell burning from stars of red clump  with He-core burning \citep{Bedding2011}. Unfortunately, the stars in the current study are not in the Kepler survey field.  However, secondary calibration based on LAMOST and APOGEE spectra of stars that have Kepler asteroseismic data and which are classified as either red clump or RGB stars \citep{Ting2018} is found to be an useful tool. Oscillations in red giants vary depending on core density. The average period spacing between oscillations of gravity mode (g- mode) and acoustic pressure mode ($p-mode$), and the frequency separation ($\Delta \nu$) between $p-modes$ are found to be key parameters to separate RC stars of He-core burning from those of H-shell burning RGB stars. 
Stars with $\Delta P \geq $ 150 and $\Delta \nu$ $\leq$ 5 have been classified as RC stars \citep{Bedding2011, Vrard2016}. The criterion is often termed as gold standard for segregating RC and RGB stars. \citep{Ting2018} estimated $\Delta P$ = 275 and $\Delta \nu$ = 4.04 for HD~24960 and $\Delta P$ = 249.3 and $\Delta \nu$ = 4.18 for TYC~1751-1713-1 suggesting that both the stars belong to the RC. Using the estimated seismic parameters ($\Delta P$, $\Delta \nu$), other stellar parameters such as $\nu_{max}$, mass and Radius of the stars (See Table~\ref{tab:asteroseismic}) have been derived using equations given in \citet{Bedding1995}. We found M = 1.27~M$_{\odot}$ for HD~24960 and M = 1.26~M$_{\odot}$ for TYC~1751-1713-1 which are very close to the values obtained using the stellar evolutionary tracks. 

Further, one can  also infer stars evolutionary phase based on the relative level of changes in C and N abundances. RC stars that have gone through the RGB phase are expected to have lower C compared to stars which are still on the RGB phase.  For example ratios of [C/N] versus [Fe/H] can be used to separate RC stars from RGB stars \citep[see][]{Masseron2017, Hawkins2016}. In Table~\ref{tab:cn}, we have given values of [C/N] and [Fe/H] of our two stars along with seven other known Li-rich giants, of which evolutionary phase for six stars is known based on asteroseismic analysis, and  based on CMD for one in open cluster Trumpler~5. We have shown all of them  in a plot (see Figure~\ref{fig:cn}) of [C/N] versus [Fe/H] of asteroseismically classified normal RC and RGB stars. The two stars in our study with [C/N]  $\leq - 0.25$ show that they are indeed RC stars. 

\begin{table}
\caption{Asteroseismic parameters derived from spectral calibration. Period spacing and frequency separation are directly adopted from \citep{Ting2018}.  Mass, radius, log~g and luminosity have been derived using scaling relations.}
\label{tab:asteroseismic}
\centering
\begin{tabular}{p{2cm} p{2cm} p{3cm}}
\hline
                       & HD~24960 & TYC1751-1713-1 \\ \hline
 $\Delta P$ [sec]      & 275.99   & 249.31 \\
 $\Delta \nu [\mu Hz]$ & 4.04     & 4.18 \\
 $\nu_{max} [\mu Hz]$  & 33.86    & 35.36 \\
 M [M$_{\odot}$]       & 1.27    & 1.26 \\
 R [R$_{\odot}$]       & 11.25    & 10.96 \\
 log~g                 & 2.44     & 2.46 \\
 $\log(\frac{L}{L_{\odot}})$ & 1.80 & 1.78\\
 \hline
\end{tabular}
\end{table}

\section{Discussion}
Derived  abundances of Li, ratios of [C/N] and $^{12}C/^{13}$C of two stars in this study, HD~24960 and TYC~1751-1713-1,  along with six other known RC giants  are summarized in Table~\ref{tab:cn}. All of them have three features in common: a) Large Li abundance which is about 1 or 2 orders of magnitude more than the maximum predicated value of A(Li)$=$1.5~dex on RGB \citep{Iben1967a, Lagarde2012}, and 2 to 3 orders of magnitude more than the generally observed values in post bump stars \citep{Lind2009b},  b) low $^{12}C/^{13}C$ ratios which are a factor of 2 to 3 lower (except for KIC~4937011, see table~\ref{tab:cn}) than the predicted values suggesting some kind of extra mixing post 1st dredge-up episode, c) low [C/N] values ( $\leq -  0.25$), indicating stars have evolved through the RGB tip  and are in RC phase\citep{Hawkins2018}. However, Li-rich giant, KIC~2305930, though it is a bonafide RC star based on asteroseismic data,  has [C/N] value which is positive  falling in the category of RGB stars (See Figure~\ref{fig:cn}). In their paper, \cite{kumar2018} commented that because of its large  $vsini$=12.5 km s$^{-1}$, and hence relatively broader spectral lines, they couldn't derive C and N abundances.  They have adopted C and N abundances given in APOGEE DR13 catalog which are based on infrared spectra of  resolution R$\approx$ 22,000. So, [C/N] value of this star should be treated with caution. 

With the recent findings of new super Li-rich red clump stars, a question arises whether there is a single mechanism 
that is responsible  for all the Li-rich RGB stars,  or there are multiple mechanisms for Li enhancement depending on their evolutionary phase on RGB. To answer this question it is necessary to have evolutionary phase unambiguously determined. Presently, there are just 7 giants which are reported as Li-rich for which evolutionary phase is determined based on asteroseismic analysis (see Table~\ref{tab:cn}). Of which 
giant KIC~9821612 reported by \citet{Jofre2015} is the only RGB Li-rich giant. Also, KIC~9821612 stands out from other Li-rich giants 
in Table~\ref{tab:cn} by having least amount of Li abundance A(Li)=1.80~dex. Recent analysis by \citep{Takeda2017} also gives similar mean abundance of
A(Li) = 1.76~dex. With the commonly applied criterion of A(Li) $>$ 1.5~dex, KIC~9821612 can be qualified as a Li-rich giant. However, as 
mentioned in \citet{Jofre2015}, its abundance is at the limit of normal Li-rich giants of A(Li) $\leq$ 1.8~dex if we take into account recent studies \citep[e.g.][]{Ruchti2011, Liu2014}. Given its relatively higher value of $^{12}C/^{13}C$ coupled with its position in HR diagram, below the bump, it is important to understand whether this star is genuinely Li-rich i.e its photosphere is enriched with Li while star is on RGB either by external or through in-situ mechanisms. Also, one can't rule out the possibility of insufficient dilution of Li from its initial value.  On other hand, its reported overabundances of $\alpha$- and $r$-process elements seem to be in odds with its RGB evolutionary phase. A more detailed study of this star is necessary to understand its Li-rich classification. 
\begin{table*}
\caption{{\bf [ }Summary of key parameters of all the known red clump Li-rich K giants including the two from this study for which evolutionary phase is determined either from asteroseismology or secondary calibration based on LAMOST spectra and
asteroseismology data.}
\centering
\label{tab:cn}
\begin{tabular}{p{3.2cm} p{1.6cm}p{1.5cm}p{0.6cm}p{0.9cm}p{1.1cm}p{0.5cm}p{2cm}p{3.6cm}}
\hline
Star name & [C/N] & [Fe/H] & A(Li) & 12C/13C& vsini & status & method  & reference\\ [1ex]
\hline
Trumpler~5 3416       & -0.22            & -0.51            & 3.75 & $14\pm3$ & $2.8$         & RC & HRD               & \cite{Monaco2014} \\
KIC~9821622           & $+0.22 \pm0.09 $  & $-0.49 \pm 0.03$ & 1.80 & $18\pm.7$& $1.01\pm.7$   & RGB& Asteroseismology  & \cite{Jofre2015}\\
KIC~5000307            & $-1.70 \pm 0.18$ & $-0.29 \pm 0.05$ & 2.71 & $<20$    & $...$         & RC & Asteroseismology  & \citet{SilvaAguirre2014} \\
KIC~4937011            & $-0.23 \pm 0.08$ & $+0.05 \pm 0.12$ & 2.30 & $25\pm5$ & $8.5 \pm 1.1$ & RC & Asteroseismology  & \citet{Carlberg2015}\\
KIC~12645107           & $-0.94 \pm 0.00$ & $-0.20 \pm 0.05$ & 3.30 & $10\pm1$ & $1.5 \pm 0.5$ & RC & Asteroseismology  & \citet{kumar2018}  \\
KIC~2305930            & $+0.26 \pm 0.00$ & $-0.50 \pm 0.10$ & 3.80 & $6\pm1$  & $12.5 \pm1$   & RC & Asteroseismology  & \citet{kumar2018} \\
2MASS~19265195+0044004 & $-0.44 \pm 0.08$ & $-0.33 \pm 0.13$ & 2.94 & $ --- $  & $2.1 \pm 2.7$ & RC & Asteroseismology  & \citet{Smiljanic2018} \\ \hline
TYC~1751-1713-1        & $-0.30 \pm 0.18$ & $-0.25 \pm 0.10$ & 4.01 & $9\pm2$  & $4.4 \pm 0.8$ & RC & HRD \& [C/N]      & This work \\
HD~24960               & $-0.70 \pm 0.20$ & $-0.45 \pm 0.05$ & 4.00 & $8\pm2$  & $6.0 \pm 1.0$ & RC & HRD \& [C/N]      & This work \\
\hline
\end{tabular}

\end{table*}

There are two main hypotheses for Li-excess in RGB stars. One being the  in-situ nucleosynthesis and subsequent extra mixing, and second is external origin. The apparent concentration of a large number of Li-rich giants at luminosity bump as reported by observations \citep{Charbonnel2000, Casey2016, Ruchti2011} led to construction of many theoretical models 
around in-situ nucleosynthesis and extra-mixing at the bump. Note, extra-mixing is also responsible for observed severe depletions of Li and very low values of $^{12}C/^{13}C$ and [C/N]. However, under some special conditions models \citep[e.g.][]{Palacios2001, Denissenkov2004, Eggleton2008, Denissenkov2011, Denissenkov2012} predict high levels of Li abundances seen in Li-rich stars. Given the relatively longer stellar evolutionary time scales at the bump and the expected brief life span of enhanced Li, models seem to explain the reported observational results: a) concentration of Li-rich stars at the bump, b) very few or lack of Li-rich stars in between the bump and RGB tip.    

The above scenario may explain Li-rich  stars if they are either at the bump or just evolved off the bump.  In case of Li-rich giants at red clump, it would be difficult to explain sustaining of Li abundance at the level shown in Table~4 as stars evolve from the bump to the clump.  The continued deep convection post-bump evolution seems to rapidly deplete remaining Li from 1st dredge-up or  Li that has been produced at the bump. This has been well illustrated by observations \citep[e.g.][]{Lind2009a}. Also, relatively much shorter Li depletion time scales compared to stellar evolutionary time scales from the bump to red clump  \citep[e.g.][]{Kumar2015} add to the argument in favour that the enhanced Li at the bump is very unlikely to  survive all the way to red clump. 
 
Alternately, accretion of Li-rich material from  sub-stellar objects such as planets or brown dwarfs may enhance photospheric Li abundance \citep{Alexander1967}. Since planet engulfment can happen anywhere along the RGB a single mechanism may be sufficient to explain reported Li-rich stars along RGB: sub-giants \citep{Li2018}, below the bump \citep{Adamow2014, Casey2016}, at the bump \citep{Charbonnel2000}, between the bump and the RGB tip, even at the red clump. External scenarios explain relatively high occurrence of Li-rich stars either at bump or at the clump may be due to relatively longer evolutionary time scales at the respective phases.

Though, at the outset, the external hypothesis seems to explain Li excess, there are genuine concerns on its ability to explain the level of Li excess observed in stars by Li-rich material accretion alone. For example \citet{AguileraGomez2016} shows a maximum limit of A(Li) = 2.2~dex in RGB stars by way of accretion of sub-stellar companion of mass lower than 15M$_{j}$ in a regime where extra-mixing does not operate. Above this mass limit, Li is expected to be depleted in the interiors of sub-stellar objects \citep{Chabrier1996}. Also, \citet{Siess1999} computations showed enhancement of photospheric Li abundance by mixing Li rich material of sub stellar objects of masses as high as 0.1M$_{\odot}$ to meet observed levels of Li.  Resulst from these studies, in the absence of induced mixing, imply that multiple number of planets and/or sub-stellar objects are required to reach the level of Li abundances observed in super Li-rich giants. However, one can't rule out the possibility of merger induced extra-mixing that brings up Li produced in H-burning shell.
Yet, we don't have evidence for such large accretion of external material in the form of enhanced metal abundances such as Fe-peak elements. One of the key evidences for the addition of external Li rich material is the enhanced abundance of $^{6}$Li which has not been observed either in RGB stars or in main sequence dwarfs with planets \citep{Reddy2002}. 
Also, models based on external events predict infrared excess as well as enhanced rotation as a consequence of accretion process which do not seem to be
evident in many of the Li-rich stars \citep{Kumar2015, Rebull2015}. Of course, IR excess may not be a stringent criterion as Li and dust shells evolve on different time scales  \citep{Kumar2015}.  As per $vsini$ values, as shown in Table~\ref{tab:cn} among  9 L-rich giants at least 4 stars do show projected $vsini$  which are two to three times of the average observed values for K giants \citep[e.g.][]{deMadeiros1996}. 

In case of two stars in this study and other RC giants in Table~4, their very large Li abundance  and relatively shorter Li depletion time scales suggest that the Li enhancement event occurred very recently either by material accretion or by in-situ nucleosynthesis, and subsequent mixing. 
In case of in-situ, He-flash at the RGB tip could be an alternate site to the luminosity bump  for excess Li seen in RC stars. He-flash at RGB tip, an immediate preceding phase of red clump phase, is a significant event on RGB at which He ignition at the core begins. There are couple of  studies that dealt with nucleosynthesis during He-flash and its effects on stellar photosphere abundances. For example, \citet{Mocak2011} show   {\bf show} injection of hydrogen into He-layers  and mixing up of interior material with outer layers resulting in very low $^{12}C/^{13}C$, large Li abundances, and enhanced N. Models also predict enhanced C as well. In case of external scenario, one can think of merger events closer to the tip.  \citet{Zhang2013} explored mergers of He white dwarfs with He-core of RGB stars to explain Li in early AGB stars, and predict enhanced Li and lower $^{12}C/^{13}C$ values along with infrared excess for certain combination of masses of He-WD and central He-cores.     
         
\section{Conclusion}
We discovered two new super Li-rich K giants with A(Li) $\approx$ 4.0~dex and argued their evolutionary phase as red clump using secondary calibration based on asteroseismic and spectroscopic data, and Gaia astrometry. The two stars add to the growing list of Li-rich red clump giants.   Large Li abundances coupled with  their RC evolutionary phase and shorter Li depletion imply that the  Li enhancement occurred very recently either by internal nucleosynthesis during He-flash at the RGB tip or by external events such as merger of sub-stellar objects near the tip.  Including two stars in this study there are 8  Li-rich K giants for which RC evolutionary phase is firmly established. Out of which five are super Li-rich stars with A(Li) $\geq$ 3.2~dex including two from this study. KIC~9821622 is the only Li-rich RGB star reported whose evolutionary phase is based on asteroseismology.  Though the numbers are too small to deduce firm conclusion, it would be interesting to comment on the large difference in number of known Li-rich giants among RC and RGB phases. Whether the difference in numbers is due to large difference in stellar evolutionary time scales between the two phases; red clump evolutionary time scales are relatively much longer compared to the bump time scales \citep{Kumar2015}.  A systematic survey of Li among the known RGB and RC giants is warranted to find clues for the origin of Li-excess. A larger survey in this direction is under progress, and will be published elsewhere. 

\section{acknowledgement}
We thank anonymous referee for his suggestions which improved the paper. 
Funding for LAMOST (www.lamost.org) has been provided by the Chinese NDRC. LAMOST is operated and managed by the National Astronomical Observatories, CAS.

\bibliographystyle{mnras}
\bibliography{ref}

\bsp
\label{lastpage}
\clearpage

\end{document}